\documentclass[12pt,preprint]{aastex}
\shorttitle{Stellar Wind Accretion in GX301-2}
\shortauthors{Leahy and Kostka}

\begin{document}

\title{Stellar Wind Accretion in GX301-2:
 Evidence for a High-density Stream}
%{E-mail:leahy@ucalgary.ca} 

\author{D.A. Leahy and M.Kostka}
\affil{Department of Physics \& Astronomy, University of Calgary,
    Calgary, AB, Canada, T2N 1N4}
%\date{Accepted  . Received 2007  ; in original form  }

%\pagerange{\pageref{firstpage}--\pageref{lastpage}} \pubyear{2002}

%\maketitle

%\label{firstpage}

\begin{abstract}
The X-ray binary system GX301-2 consists of a neutron star in an 
eccentric orbit accreting from the massive early-type star WRAY 977. It has
previously been shown that the X-ray orbital light curve is consistent with
existence of a gas stream flowing out from WRAY 977 in addition to 
its strong stellar wind. Here, X-ray monitoring observations by the 
Rossi X-ray Timing Explorer (RXTE)/ All-Sky-Monitor (ASM) and pointed observations
by the RXTE/ Proportional Counter Array (PCA) over the past decade are analyzed.
We analyze both the flux and column
density dependence on orbital phase. The wind and stream dynamics are calculated
for various system inclinations, companion rotation rates and wind velocities, as
well as parametrized by the stream width and density. These calculations are used
as inputs to determine both the
expected accretion luminosity and the column density along the line-of-sight to the
neutron star. The model luminosity and column density are compared to observed flux
and column density vs. orbital phase, to constrain the properties of the stellar
wind and the gas stream. We find that the change between bright and medium intensity
levels is primarily due to decreased mass loss in the stellar wind, but the change 
between medium and dim intensity levels is primarily due to decreased stream density.
The mass-loss rate in the stream exceeds that in the stellar
wind by a factor of $\sim$2.5.
The quality of the model fits is significantly better for lower inclinations, 
favoring a mass for WRAY 977 of $\sim53-62 M_{\odot}$.
\end{abstract}


\section{Introduction}

GX 301-2 (also known as 4U 1223-62) is  a pulsar with a 680 s rotation period, 
in a 41.5 day eccentric orbit  \citep{b32}. 
The mass function is $31.8 M_{\rm  \odot}$, making
the minimum companion mass $35 M_{\rm  \odot}$ for a $1.4 
M_{\rm  \odot}$ neutron star. 
The companion Wray 977 has a B1 Ia+ spectral classification  \citep{b9}, 
determined via comparison with the hypergiant $\zeta^{1}$ Sco.  
This analysis also yielded an upper limit for the radius of Wray 977 
of $75  R_{\rm  \odot}$ placing it just inside its tidal radius.

The neutron star flares regularly in X-rays approximately 1-2 days before 
periastron passage, and several stellar wind
accretion models have been proposed to explain the magnitude of the 
flares and their orbital phase dependence 
%(e.g.  \citep{koh97}, \citep{lea91}, \citep{hab91}).
% need to spell out references with non-standard format e.g. here:
(e.g. Koh et al., 1997, Leahy, 1991, Haberl, 1991).
The modeling by \cite{b18} and \cite{b5} was done using TENMA and EXOSAT
observations, respectively, which cover many short data sets spaced irregularly
over orbital phase.
More recently better orbital phase coverage has been obtained by CGRO/BATSE
 \citep{b13}, 
which however has much lower sensitivity than the previous studies.

The broad band X-ray spectrum of GX301-2 has been studied by TENMA (Leahy \& Matsuoka, 
1990, Leahy \& Matsuoka, 1989a and Leahy \& Matsuoka, 1989b) and ASCA measurements
\citep{b31}. The latter study illustrates the complexity of the GX301-2
spectrum. Four components are necessary: 
i) an absorbed  power law with high column density;
ii) a scattered  power law with much lower column density; iii) a thermal
component with temperature of 0.8 keV; iv) a set of six emission lines 
(including the iron line at 6.4 keV).   Of the above, components
ii) and iv) are due to
reprocessing in the gas in stellar wind from Wray 977 which surrounds the x-ray
source. Reprocessed spectra were calculated for a centrally illuminated cloud 
by \citep{b17} and for an externally illuminated cloud by \citep{b16},
including the Comptonized iron line shapes. Later, the Comptonized iron line was
detected in GX301-2 using the Chandra High Energy Transmission Grating \citep{b36}, 
confirming the high column density (on the order of $10^{24}$ cm$^{-2}$) and 
yielding an upper limit on electron temperature of $\sim$3 eV.
The orbital phase dependence of the X-ray spectrum of GX301-2 was observed by 
the PCA on board RXTE  \citep{b25}.  It was concluded that clumpiness in the matter 
surrounding the neutron star caused large variability in column density measurements.

Long-term monitoring of GX301-2 has been carried out
by the ASM on board RXTE.  Analysis of these observations for the ~5 year time period 
MJD\,50087.2 to MJD\,52284.5 was done by \citep{b15} (henceforth refered to as L02).
In this paper the light curve based on the significantly longer ~10 year RXTE/ASM 
database is analyzed.  In addition we study the flux and column density
measurements made by the RXTE/PCA, as well as column densities derived from the 
RXTE/ASM softness ratios.  Improved modeling methods are introduced: accurate analytic 
description of the stream and inclusion of simultaneous flux and column density 
calculations.  The inclusion of this much more extensive data and the more realistic 
modeling results in significant improvement on constraints on the system properties 
of GX301-2 and allows new conclusions to be drawn. 

\section[]{RXTE/ ASM and RXTE/ PCA Observations}

  The ASM on RXTE  \citep{b23} 
consists of three scanning shadow cameras (SSC's), 
each with a field of view of $6^\circ$ by $90^\circ$ FWHM. The SSC's are rotated
in a sequence of ``dwells'' with an exposure typically of 90 seconds,
 so that the most of the sky can be covered in one day. The dwell data are
also averaged for each day to yield a daily-average.
The RXTE/ ASM dwell data and daily-average data were obtained from the
ASM web site.
%(http://xte.mit.edu/ASM_lc.html). 
The data reduction to 
obtain the count rates and errors from the satellite observations was
carried out by ASM/RXTE team, and the procedures are described at
the web site. 
The ASM count rates used here include the full energy range band as well as
the three sub-bands 1.3-3.0 keV, 3.0-5.0 keV and 5.0-12.1 keV. The data covered
the time period MJD\,50172.6 to MJD\,53978.6. The regular outbursts every 41.5 day 
orbital cycle are seen in the 5-12 keV count rates, as well as the variability 
from cycle to cycle.

The orbital parameters of GX301-2, updated with the BATSE observations
\citep{b13} and used for the current study are as follows.
$P_{\rm orb}=41.498$days, $a_x sin(i)=368.3$lt-s, eccentricity $e=0.462$,
longitude of periastron $\omega=310^\circ$, time of 
periastron passage $T_0=$MJD\,48802.79. 

The dwell data is used in the analysis that follows.
The three RXTE/ ASM  sub-bands and full energy range band were epoch folded. 
The orbital light curves for these bands and the full energy range 
band are shown in Fig. 1 with orbital phase zero defined by the
time of periastron passage, $T_0$.

GX301-2 shows a significant variability above statistical uncertainties in 
intensity from orbit to orbit.
This is illustrated in Fig. 2 which shows the RXTE/ ASM data over the entire observation 
period with timebins equal to one orbital period.  
The r.m.s. variability is 0.33 ASM c/s compared to a mean error of 0.044 ASM c/s: the
variability is real at greater than 7 sigma significance.
There is a secular decrease in the
mean flux  in the amount of -0.07 ASM c/s/year. 
However the length of the data set is not long enough to establish a long term trend,
and the flux is also consistent with no secular decrease after $\sim$MJD\,51200.
The high time-resolution data were tested for long term periodicities 
by examining  $\chi^2$ vs. period for
epoch folding over periods up to 500 days. This showed peaks at N times the 
orbital period (with N=1, 2, 3, 4 ...): this is due to aliasing of the orbital light 
curve. To negate the effect of aliasing, one bin per orbital period was used as input 
to the epoch  folding. This then yielded peaks at 4 and 8 times the orbital period.  
A visual inspection of Figure 2 verifies that this long-term period is real: there 
are prominent oscillations around MJD 52000 which have a period $\sim$ 4 times 
the orbital period.   

Fig. 3 shows the 5-12.1 keV orbital light curves when the the total time period is 
divided into three different intensity levels based on the average count rate per 
orbit: bright (average count rate per orbit greater than 2 c/s); medium (average 
count rate per orbit in range 1.5- 2 c/s) and dim (average count rate per orbit 
less than 1.5 c/s). 
The variability in the shape of the light curve for GX301-2 between bright, medium 
and dim levels is primarily due to variability in
the outburst peak near orbital phase 0.9.   The medium and dim level folded light
curves are consistent with each other between orbital phases 0.1 to 0.8, and the bright 
level light curve is different from medium and dim between
orbital phases 0.3 to 0.55.

Column densities were extracted from the RXTE/PCA spectral fits of \cite{b25}. The 
values used were the column densities of the absorbed component, since that measures 
the column density to the neutron star. Whereas the column densities of the 
scattered component are complicated to interpret and represent mean values to 
the scattering region. The orbital phase coverage of the RXTE/PCA column densities 
is not very uniform, and all column densities are from a single orbit observation 
of GX301-2. To obtain better orbital coverage and to cover the same multiyear timespan 
as the ASM lightcurve observations, an estimate of column density versus orbital phase
 was created based on the 3-5 keV to 5-12 keV softness ratio of the ASM observations.  
Conversion coefficients from the softness
ratio to column density were determined using NASA's WebPIMMS software assuming a 
power law spectrum with photon index -1.0. Figure 4 shows the derived ASM column 
densities compared to the observed PCA column densities. 
The main approximation in calculation of column densities from ASM softness ratio is 
use of a single power law spectrum, which is the equivalent of ignoring the 
scattering contribution to the spectrum.  

\section{Model}

\subsection{Wind Model}

The stellar wind velocity and density profiles are considered first.
Both radial and azimuthal velocity components of the wind were included in 
this analysis. 
The radial wind velocity follows a power law and is taken to be of the form, 
$v_w(r)=v_o (1-R_s/r)^{\beta}+c_s$ with $\beta=1$, $c_s$ the speed of sound and 
$v_o$ the terminal velocity of the wind,  \citep{b3}.  Conservation of angular 
momentum dictates that the azimuthal component of the wind velocity drops off as $1/r$.  
The constant stellar angular speed $\omega$ of WRAY977 is expressed using the 
parameter $f$:
\begin{equation}
\omega(f)=f\times \omega_{\rm orb} +(1-f)\times \omega_{\rm per}
\end{equation} 
with 
$\omega_{\rm orb}$ is
the average orbital angular velocity ($2\pi/P_{\rm orb}$) and 
$\omega_{\rm per}=\omega_{\rm orb}(1+e)^{0.5}/(1-e)^{1.5}= 3.06\omega_{\rm orb}$ is
the periastron angular velocity. Thus the primary is taken to be rotating at
some angular velocity between $\omega_{\rm orb}$ and $\omega_{\rm per}$. The large 
difference in  $\omega_{\rm orb}$ and $\omega_{\rm per}$ is due to the high eccentricity
of the orbit. 
Fits of a wind driven accretion model for GX301-2 were studied by L02 and found to 
be unable to fit the "double bump" nature of the light curve.  As well L02 tested 
a wind and disk model of accretion for GX301-2 which could not fit the observations.
  The conclusion drawn by L02 was that a wind and stream model \citep{b35} is the 
best supported model for GX301-2.

\subsection{Wind plus stream model}

Simplified models describing a simultaneous wind and stream accretion process were 
first used by \cite{b5} (straight line stream) and \cite{b18} (spiral stream)
 to fit the less complete data from EXOSAT and TENMA. 
In the model \citep{b35}, a stream originates at the point on the surface of 
WRAY 977 that is nearest to GX301-2.  The stream then bends backwards (with respect 
to the direction of orbital velocity of GX301-2) as it travels radially outward from 
the primary star. 

Here we calculate the stream position at any given orbital phase by integrating 
the radial and azimuthal equations of motion.  Roughly speaking the stream is like 
an Archimedes spiral co-rotating at the orbital angular velocity.  However since 
the point of origin of the stream follows the neutron star it has a greatly varying 
angular velocity, by a factor of $[(1+e)/(1-e)]^2=7.4$ for GX301-2.  The result is 
a stream that changes shape considerably with orbital phase, similar to a 
garden sprinkler with an uneven rotational speed. 
Animations of the stream can be found at www.iras.ucalgary.ca/$\sim$leahy/.  The 
stream shape depends on the terminal wind velocity ($v_o$), the angular speed 
($\omega(f)$) of WRAY 977, the system inclination, and on companion radius (through 
the wind law).  The model light curve depends also on the stream width and density 
and the speed of sound.  In order to calculate the full stream shape we started 
with a set of terminal wind velocities and stellar angular velocities 
($f$) then created a stream for each combination of $v_o$ and $f$.  
To allow $v_o$ and $f$ to be free parameters, the stream position was interpolated in 
$f$ and $v_o$.      

Analysis done by L02 suggests that two crossing points (with orbital phases 
$\gamma_{1}, \gamma_{2}$) between GX301-2 and the stream exist.  While L02 allowed 
$\gamma_{1}, \gamma_{2}$ to be free parameters, here we constrain them to be 
governed by the computed stream shape.  The relative velocity of GX301-2 with respect 
to the stream and the stream density both play a large role in the intensity 
of the X-ray flux. 
%We note that the Archimedes spiral-like stream rotates once per system orbit, but due to its strongly changing shape  
One stream crossing occurs just before periastron (orbital phase $\sim$0.93). GX301-2 
is nearing its peak speed to overtake the stream, but the stream is near its 
highest density and lowest radial velocity resulting in a large increase in luminosity. 
The second stream crossing is at orbital phase $\sim$0.55. As GX301-2 approaches 
apastron it slows to its most leisurely pace, so the stream is able to overtake the 
neutron star. Since the radial wind speed is highest and stream density lowest at 
apastron, a significantly lower peak in luminosity occurs.  
Physically one expects the stream width to increase with radial distance from the
companion star, due to expansion of the higher density, overpressured stream in the
lower density surrounding stellar wind. Here we use a Gaussian density profile with
variable width. The angular width ($\sigma$) is taken as a power law function 
of distance from the center of the companion star: 
$\sigma (r)= \sigma_0 (r/r_{per})^{-\kappa}$, with $\sigma_0$ the width normalized at
periastron distance $r_{per}$.
$\kappa<0$ corresponds to a stream with increasing physical width
as the stream propagates outward. 
We ensure mass conservation by employing the continuity equation, so the density of
the stream varies with $r$ depending on the value of $\kappa$.
The angular width of the stream (viewed from the companion star) 
depends on $(r/r_{per})^{-\kappa-1}$, so $\kappa>-1$ corresponds to a stream with 
decreasing angular width as the stream propagates outward. 

\subsection{Comparison to Data}

GX301-2 has a significant absorption by its stellar wind in soft X-rays
with column densities several times $10^{23}$ cm$^{-2}$. This shows up well 
in Fig. 1 above:  
Band 1 and Band 2 data are affected significantly by absorption but Band 3 
(5-12.1 keV) is mostly free of absorption effects since the photoelectric 
cross-section is very small above 3-4 keV. 
This is also confirmed by the consistency in shape of the Band 3 light curve with
the BATSE light curve \citep{b13}, although the RXTE/ ASM light curve here is
of significantly better statistical quality.
Thus the  5-12.1 keV band flux is
taken here to be proportional to the X-ray luminosity of the pulsar.

The comparison of our wind plus stream models to the RXTE/ ASM orbital light
curve is made by $\chi^2$ minimization using the non-linear
conjugate gradient method. For each minimization (i.e. fitting), some parameters 
were taken as fixed parameters (stellar radius and system inclination) and the
remaining parameters were taken as free parameters.

The mass-radius constraints on WRAY 977 were discussed in detail in L02. Here
we replot the radius contraints in Fig. 5. We use inclination rather than mass as the 
independent variable, since radius and inclination are critical inputs to our
model calculations. Also we have extended the upper limit of $T_{eff}$ to 22500 K
as suggested by the data of \cite{b9},
and show the most relevant region of the radius vs. inclination diagram. Allowable
companion radii are smaller than those that result in eclipses (i.e. left of the 
"Eclipse Limit" line) and smaller than those that overflow the Roche lobe (below
the "Mean Roche Radius" line). These are hard constraints that cannot be violated.
The estimated mass-radius relation and estimated $T_{eff}$ give that the star should
fall in between the $T_{eff}$=20000 K and 22500 K lines. 
Thus for our models, we chose allowable inclination and radius pairs falling in
the allowed region from Fig. 5. We find a maximum radius of 78 $R_{\odot}$ consistent
with the maximum of \citep{b9}, and a minimum radius of 45 $R_{\odot}$. 
These correspond to inclinations of 52$^\circ$ and 77.5$^\circ$.

Different pairs of stellar radii and inclination were chosen covering the allowed 
region in the radius-inclination plane.  These are listed in the first two colums
of Table 1. 
For each case of the two fixed parameters; R$_c$ and inclination,  
the stream model was fit to the ASM data.  
The free parameters in the model were terminal wind velocity ($v_o$), 
stream density contrast (ds), stream angular width in radians ($\sigma_o$), stream
width variation parameter $\kappa$,  mass-loss rate ($\dot{m}$), 
a normalization factor, and the stellar 
angular velocity factor ($f$) that determined the rate of rotation of WRAY 977.
Initial fits were done on the light curves and column densities from the full 
ASM data set, then fits were
carried out on the three subsets of lightcurve and column density data
for the different intensity levels
(bright, medium and dim). Since the light curves for the three different 
intensity levels 
were significantly different, we present results for separately fitting the 
different intensity levels. 
We carried some joint fits to both lightcurve and column density, and found no 
significant difference to fits done to lightcurve to determine all model lightcurve 
parameters, followed by fits to column density to determine wind mass loss rate.
This is not too surprising as the errors on lightcurve data are very small, whereas
the errors on the column density data are large (see Fig.3 and Fig.4).
Thus the the light curve completely dominated the  determination of joint parameters.
The second procedure has the two advantages of: much faster run-time; and 
less reliance on the approximate column densities.  

\section{Results}

For each set of fixed parameters (radius,$R_{c}$, and inclination $inc)$ in Table 1), 
the best fit parameters for fitting to the ASM light curve and column density data
 are listed in Table 1. B, M and D refer to bright, medium and dim intensity levels.
The fit parameters for the light curve fits are base wind velocity ($v_{wo}$), 
stellar angular velocity parameter ($f1$), stream width and width 
variation parameters ($\sigma_{o}$ and $\kappa$), and stream 
central density enhancement at periastron ($ds$).
The $\chi_L^2$ values for the light curve fits are listed in column 8.
The large $\chi_L^2$ compared to the number of degrees of freedom (34) 
shows that the model does not provide a statistically good fit. 
Fig. 6 shows model fits to the ASM lightcurve data (top panel) and ASM column 
density data (bottom panel) for bright level for 
$R_c=75R_{\odot}$, inclination 55$^\circ$. The shape of the light curve is fit
very well: the contributions to $\chi_L^2$ mainly come from real fluctuations in
the observed light curve at orbital phases $\sim$0.3- 0.4 and $\sim$0.6 - 0.7. 
This is likely due to clumps in the stellar wind and stream that we cannot model
currently: \cite{b25} also noted large fluctuations in their RXTE/PCA observations
which they attributed to clumps in the stellar wind.
%Similarly, the fit to shape of the column density vs. orbital phase is good and the
%main contributions to $\chi_N^2$ come mainly from fluctuations from a smooth variation.

There are several trends in the fit results. As one goes to smaller $R_c$ and larger
inclination, the best fit wind velocity decreases: from $\sim$600 km/s to 
$\sim$300 km/s. 
The derived windspeeds are in good agreement with the values estimated from the
optical spectrum of WRAY 977 (\cite{b9}).
In all cases, for a given $R_c$ and inclination, the wind velocity 
is highest for bright and lowest for dim.  
Also in all cases the best fit mass-loss
rate is highest for bright and similar for medium and dim, where as the density
enhancement ratio, $ds$, is smallest for bright and highest for medium. 
Instead, one can consider the central density of the stream (proportional to 
density enhancement ratio, column 7, times mass-loss rate, column 9) 
This shows that the central density of the stream is essentially unchanged between
bright and medium levels, so the main change between bright and medium is the 
stellar wind mass-loss rate. However the central density of the stream drops from
medium to dim whereas the wind mass-loss rate does not change signficantly. 
Thus the main change between medium in dim is that the stream density is 
lower for dim. 

The angular rotation rate parameter $f1$ systematically decreases with $R_c$.
From the stream model, this is just due to the requirement of 
having the neutron star- stream crossings at
the correct observed orbital phases as the position of the base of the wind 
(at $R_c$) changes.
From equation 1, a value of $f1$=1 is for the companion rotation synchronous 
with the mean orbital rotation
and a value of 0 is for synchronous with periatron angular velocity. 
Tidal torques in the eccentric orbit would yield an intermediate value, consistent
with the best-fit values in Table 1.

Values of $\kappa$ are in the range -0.4 to 0, thus the stream 
physical width grows with $r$, as expected, and the stream angular 
width decreases with $r$ ($-1<\kappa<0$).
The stream density enhancement over that of the spherical component of the stellar
wind is typically 25 (with a range of 20 to 30), and the stream angular width
at periastron
is in the range of 22$^\circ$ to 26$^\circ$. 

The column density model gives two gradual peaks (see Fig. 6): one near periastron
and one near orbital phase 0.25. The stream is seen to dominate over the wind component
for the periastron peak but both wind and stream contribute roughly equally for the
peak near phase 0.25. The system inclination can be such that the neutron star
is nearly eclipsed by the companion (see Fig. 5 for $R_c$-inclination values where
this occurs, near the "Eclipse Limit" line). In this case the wind contribution to
the column density becomes large near orbital phase 0.25, when the line-of-sight
passes near the companion's surface, and the $\chi_N^2$ for the column density fit
becomes large. An example of this is the (48$R_{\odot}$,75$^\circ$) case in Table 1.

Finally, we have calculated the mass-loss rate in the stream and added this as 
column 10 in Table 1. It is seen that the stream mass-loss rate is about a factor
of 2 to 2.5 times higher than the mass-loss rate in the wind. This is a dramatic 
confirmation of the importance of the stream in the total mass-loss rate from
WRAY 977. It is also consistent with WRAY 977 being close to filling its Roche lobe.
Since this can only occur at lower inclinations without having WRAY 977 too close to 
eclipsing, this is another indicator that the system inclination is near the low end
of the allow range (near $\sim55^\circ$).

\section{Conclusion}

Long term (10-year) monitoring of GX301-2 with the RXTE/ASM has revealed some new
properties of this high-mass X-ray binary. Secular changes in flux (Fig. 2)
are also accompanied by flux oscillations with a period of four 41.5-day binary 
orbits. The orbital light curve is seen to be significantly different between bright,
medium and dim flux levels (Fig.3).

We have constructed an improved stellar wind and stream model for the GX301-2/ WRAY 977
binary system, extending the work of L02. 
The model is compared to long term lightcurve observations from
the RXTE/ASM and to column densities derived from the RXTE/ASM 3-5 keV to 5-12 keV
softness ratios.
We have validated the necessity of including a stream in the mass
flow from WRAY 977 in addition to a spherically symmetric wind in order to explain
the observed light curves and column densities.
The timings and amplitudes of the
two peaks in the lightcurve, near orbital phases 0.92 and 0.5 are naturally explained
by accretion onto the neutron star from an Archimedes spiral-type stream. The
quality of the column density data is low due to statistical uncertainties. 
Yet the column density data provides the primary constraint on the stellar 
wind mass loss rate. 
%and weaker independent constraints on the wind, stream and system parameters.
We have found best fit parameters for a range of radii for WRAY 977 and a range of 
system inclinations which are consistent with the physical constraints such as no
eclipse and maximum mean radius not exceeding the mean Roche radius (see Fig. 5).
From the model fits, we find the change between bright and medium intensity
levels is primarily due to decreased mass loss in the stellar wind, but the change 
between medium and dim intensity levels is primarily due to decreased stream density.
For any intensity level, the total mass-loss rate in the stream exceeds that in the 
wind by a factor of $\sim$2.5, indicating the crucial role of the stream in this
binary system.

The model fits at higher inclinations are significantly worse than those at 
lower inclinations.
In Table 1, last column, we list the total $\chi^2$ (values for lightcurve and 
column density fits summed). 
The best fit values of ($R_{c},inc$) can be determined by summing the total $\chi^2$ 
for the three intensity levels B, M and D to yield a total $\chi_{BMD}^2$. 
This gives, in order of increasing $chi_{BMD}^2$,: 
($R_{c},inc$)=(75$R_{\odot}$,$55^{o}$) with  $\chi_{BMD}^2=1660$;
(68$R_{\odot}$,$60^{o}$) with $\chi_{BMD}^2=1820$; and 
(62$R_{\odot}$,$60^{o}$) with $\chi_{BMD}^2=2010$. 
Thus the $chi_{BMD}^2$ from our modeling strongly prefers the lowest inclination.
In Table 2, we have listed
the companion mass vs. inclination for GX301-2. Thus the wind plus stream model fits
to the RXTE/ASM lightcurve and column density data favor $\sim55-60^\circ$ inclination,
or companion mass $\sim53-62 M_{\odot}$. 

\begin{table*}
 \centering
  \caption{Best Fit Parameters.}
  \begin{tabular}{@{}lllrrrrrllrr@{}}
  \hline
   	($R_{c}, inc)$ & Data & $v_{wo}$ & $f1$ & $\sigma_{o}$ & $\kappa$ & $ds$ & $\chi_{L}^2$ & $\dot{m}_{wind}$ & $\dot{m}_{stream}$ & $\chi_{N}^2$ & Total $\chi^2$ \\
 & & [km/s] & & & & & &$[M_{\odot}/yr]$ & $[M_{\odot}/yr]$ & &\\       
 \hline
   (75$R_{\odot}$,$55^{o})$ & B & 630 & 0.44 & 0.44 & -0.40 & 25 & 430 & 3.8x$10^{-5}$ & 9.2x$10^{-5}$ & 140 & 570   \\
  & M & 600 & 0.48 & 0.43 & -0.31 & 31 & 690 & 3.0x$10^{-5}$ & 8.6x$10^{-5}$ & 99 & 790 \\
 & D & 580 & 0.51 & 0.46 & -0.31 & 26 & 210 & 2.7x$10^{-5}$ & 7.4x$10^{-5}$ & 88 & 300 \\
\hline
(68$R_{\odot}$,$60^{o})$ & B & 570 & 0.38 & 0.43 & -0.32 & 22 & 460 & 3.4x$10^{-5}$ & 6.9x$10^{-5}$ & 160 & 620 \\
 & M & 540 & 0.42 & 0.42 & -0.23 & 28 & 750 & 2.6x$10^{-5}$ & 6.4x$10^{-5}$ & 130 & 880 \\
 & D & 500 & 0.45 & 0.45 & -0.23 & 24 & 230 & 2.2x$10^{-5}$ & 5.3x$10^{-5}$ & 89 & 320 \\
\hline
(62$R_{\odot}$,$70^{o})$ & B & 500 & 0.35 & 0.43 & -0.28 & 22 & 500 & 1.2x$10^{-5}$ & 2.4x$10^{-5}$ & 490 & 990 \\
& M & 470 & 0.39 & 0.42 & -0.19 & 28 & 780 & 7.3x$10^{-6}$ & 1.8x$10^{-5}$ & 550 & 1300 \\
 & D & 450 & 0.41 & 0.44 & -0.20 & 24 & 250 & 8.8x$10^{-6}$ & 2.0x$10^{-5}$ & 170 & 420 \\
\hline
(62$R_{\odot}$,$65^{o})$ & B & 510 & 0.32 & 0.42 & -0.24 & 21 & 530 & 2.6x$10^{-5}$ & 4.8x$10^{-5}$ & 220 & 750 \\
 & M & 480 & 0.39 & 0.42 & -0.15 & 26 & 820 & 1.9x$10^{-5}$ & 4.4x$10^{-5}$ & 210 & 1000 \\
 & D & 450 & 0.39 & 0.44 & -0.17 & 23 & 250 & 1.7x$10^{-5}$ & 3.8x$10^{-5}$ & 100 & 350 \\
\hline
(62$R_{\odot}$,$60^{o})$ & B & 520 & 0.29 & 0.42 & -0.19 & 20 & 570 & 3.4x$10^{-5}$ & 6.0x$10^{-5}$ & 140 & 710 \\
 & M & 480 & 0.34 & 0.41 & -0.10 & 25 & 860 & 2.5x$10^{-5}$ & 5.3x$10^{-5}$ & 100 & 960 \\
 & D & 450 & 0.36 & 0.45 & -0.13 & 22 & 260 & 2.2x$10^{-5}$ & 4.9x$10^{-5}$ & 84 & 340 \\
\hline
(55$R_{\odot}$,$70^{o})$ & B & 450 & 0.22 & 0.41 & -0.13 & 20 & 650 & 1.6x$10^{-5}$ & 2.7x$10^{-5}$ & 380 & 1000 \\
 & M & 420 & 0.28 & 0.40 & -0.05 & 25 & 930 & 9.7x$10^{-6}$ & 1.9x$10^{-5}$ & 440 & 1400 \\
 & D & 390 & 0.30 & 0.44 & -0.08 & 22 & 300 & 9.9x$10^{-6}$ & 2.1x$10^{-5}$ & 140 & 440 \\
\hline
(51$R_{\odot}$,$70^{o})$ & B & 430 & 0.15 & 0.40 & -0.05 & 19 & 740 & 1.7x$10^{-5}$ & 2.6x$10^{-5}$ & 330 & 1100 \\ 
 & M & 400 & 0.20 & 0.40 & +0.02 & 24 & 1000 & 1.1x$10^{-5}$ & 2.1x$10^{-5}$ & 390 & 1400 \\
 & D & 370 & 0.17 & 0.43 & -0.02 & 21 & 330 & 1.0x$10^{-5}$ & 1.9x$10^{-5}$ & 130 & 460 \\
\hline
(48$R_{\odot}$,$75^{o})$ & B & 410 & 0.09 & 0.39 & -0.005 & 19 & 840 & 3.4x$10^{-6}$ & 4.9x$10^{-6}$ & 710 & 1600 \\
 & M & 380 & 0.15 & 0.39 & +0.07 & 23 & 1110 & 1.6x$10^{-6}$ & 2.8x$10^{-6}$ & 740 & 1900 \\
 & D & 350 & 0.17 & 0.43 & +0.01 & 21 & 370 & 2.4x$10^{-6}$ & 4.7x$10^{-6}$ & 250 & 620 \\
\hline
\end{tabular}
\end{table*}

\begin{table*}
 \centering
  \caption{Mass and Inclination.}
  \begin{tabular}{@{}llllll@{}}
  \hline
   	$inc$ & $55^{o}$ & $60^{o}$ & $65^{o}$ & $70^{o}$ & $75^{o}$  \\
     $M_{c}$ & $62M_{\odot}$ & $53M_{\odot}$ & $47M_{\odot}$ & $42M_{\odot}$ & $39M_{\odot}$  \\      
\hline
\end{tabular}
\end{table*}

%\newpage

\section*{Acknowledgments}
DAL acknowledges support from the 
Natural Sciences and Engineering Research Council.

\clearpage


%\begin{figure}
%\plotone{f1.eps}
%\caption{This is the first figure and it uses sgi9259.eps as
%its EPS figure file. \label{fig1}}
%\end{figure}

\begin{figure}
%\vspace{302pt}
%\includegraphics[width=84mm]{fig1.eps}
\plotone{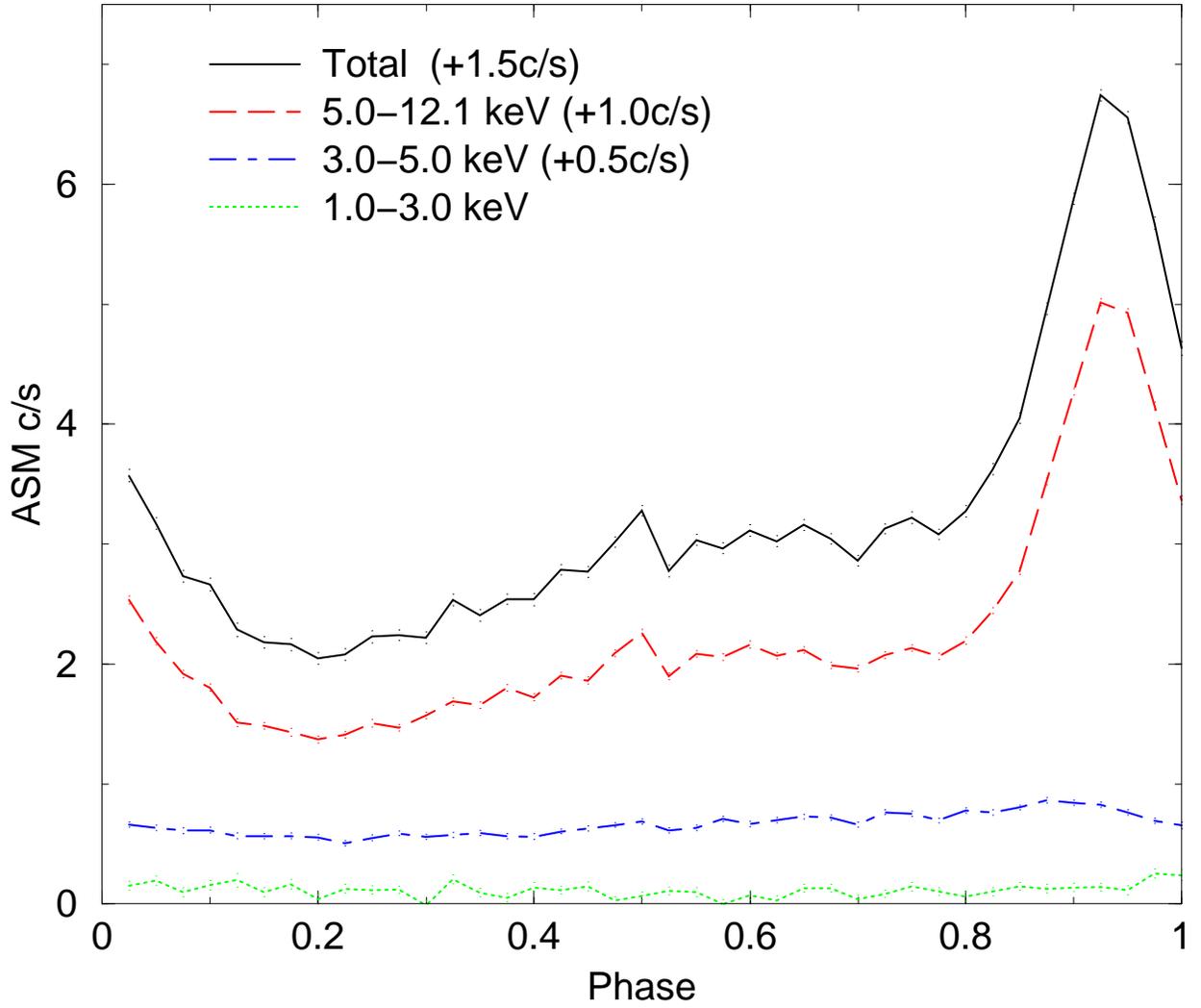}
\caption{Orbital light curves of GX301-2 in the three RXTE/ ASM energy bands and
 total energy band  from the ASM dwell data folded at 41.498 days.}
%\label{Fig. 1}
\end{figure}

\begin{figure}
%\vspace{302pt}
%\includegraphics[width=84mm]{fig2.eps}
\plotone{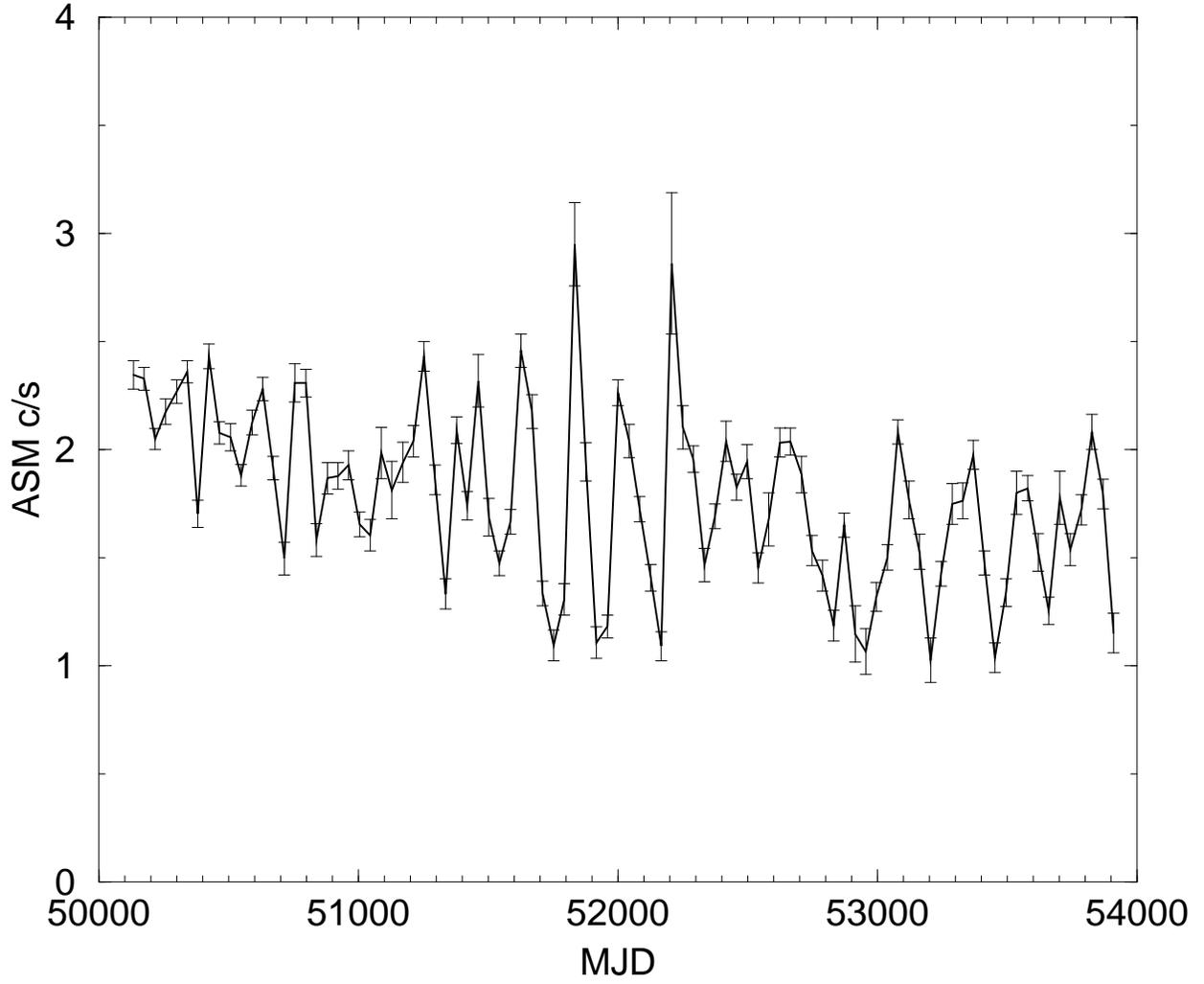}
\caption{Long term variability: RXTE/ ASM data for the entire observation 
period with timebins equal to one orbital period (41.498 days).}
%\label{Fig. 2}
\end{figure}

\begin{figure}
%\vspace{302pt}
%\includegraphics[width=84mm]{fig3.eps}
\plotone{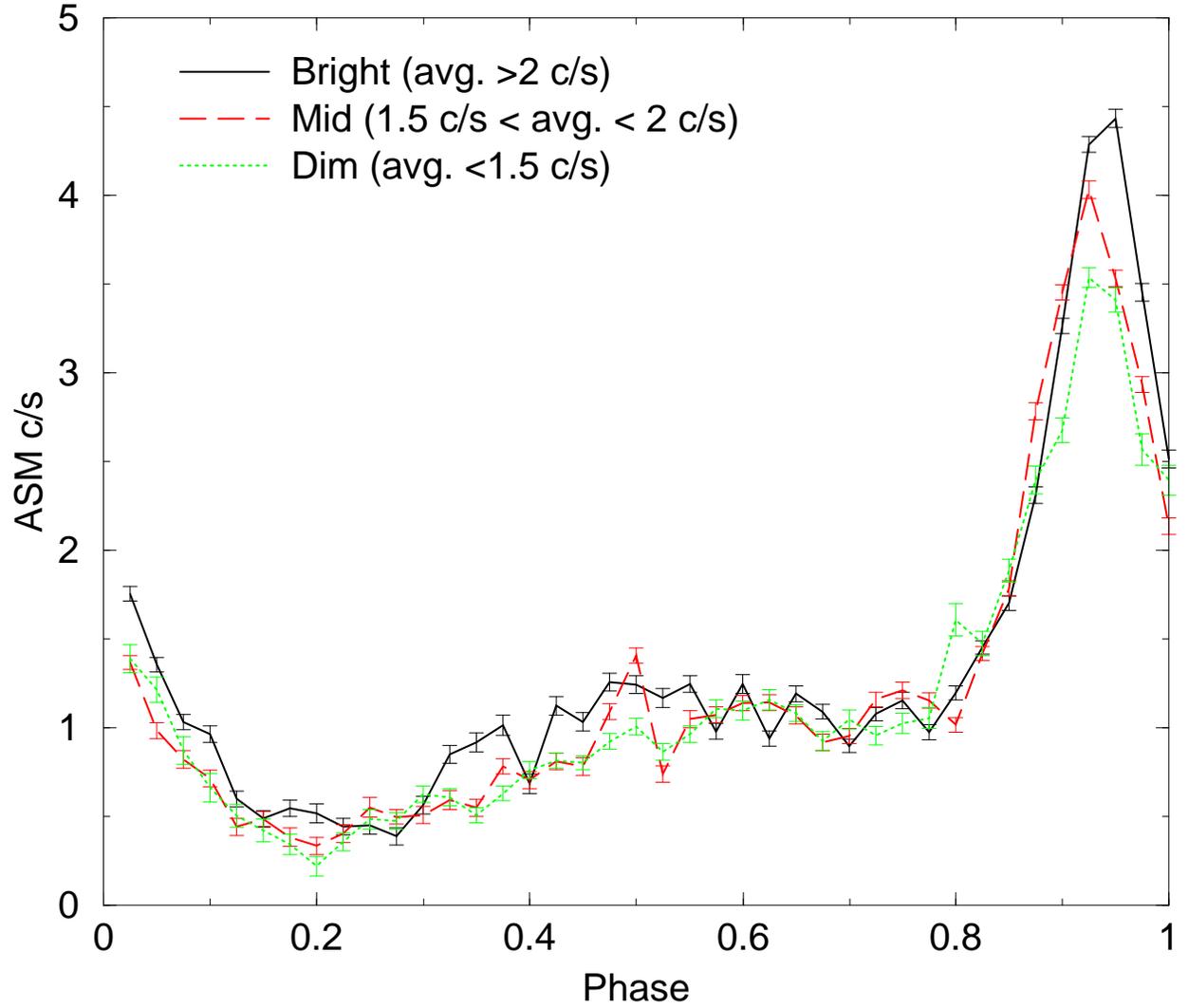}
\caption{5-12.1 keV orbital light curves for bright (average count rate per orbit 
 greater than 2 c/s); medium (average count rate per orbit in range 1.5- 2 c/s) and dim 
(average count rate per orbit less than 1.5 c/s).}
%\label{Fig. 3}
\end{figure}

\begin{figure}
%\vspace{302pt}
%\includegraphics[width=84mm]{fig4.eps}
\plotone{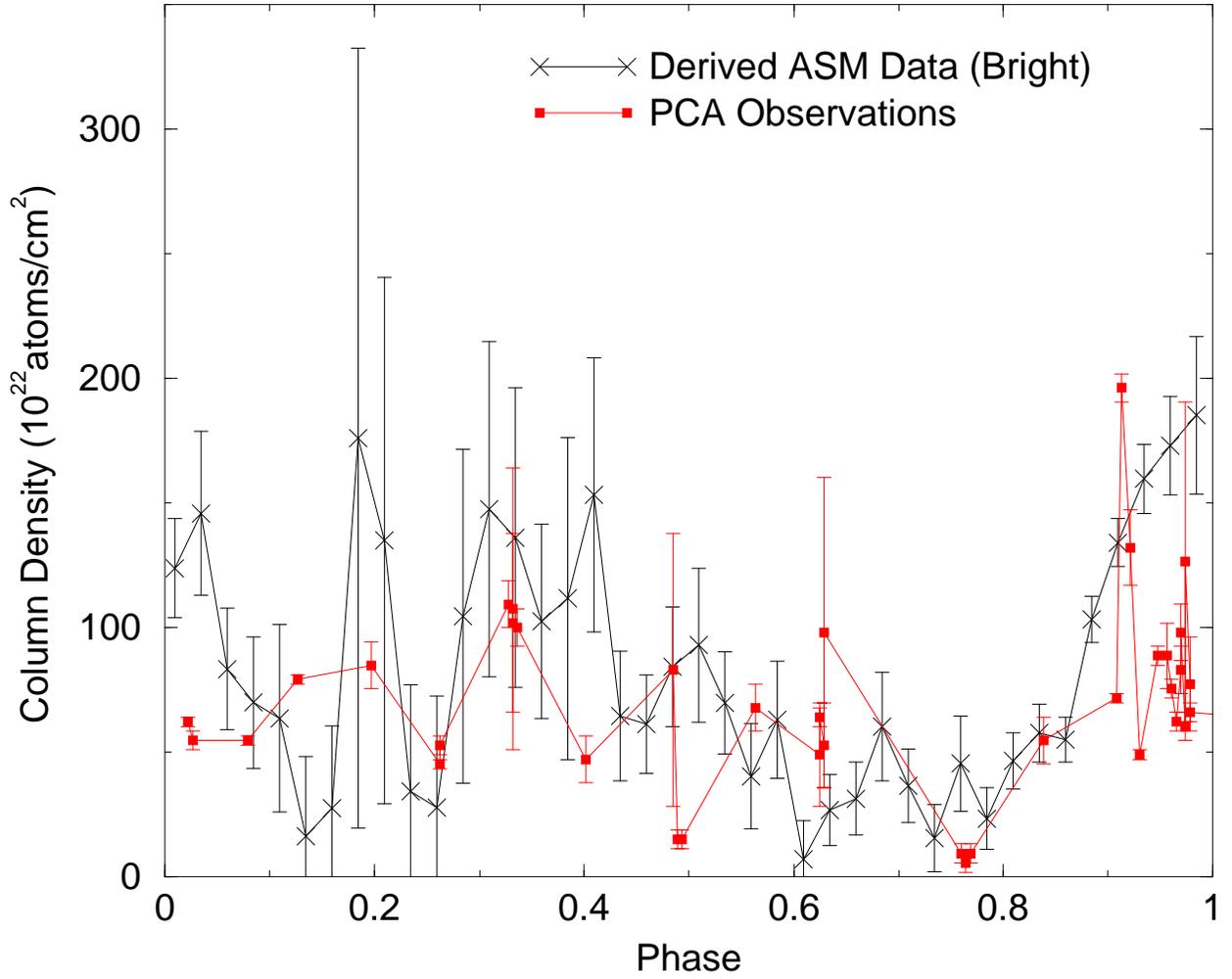}
\caption{Column densities derived from ASM 3-5 keV to 5-12 keV softness ratio for
bright level data, compared to the observed PCA column densities.}
%\label{Fig. 4}
\end{figure}

\begin{figure}
%\vspace{302pt}
%\includegraphics[width=84mm]{fig5.eps}
\plotone{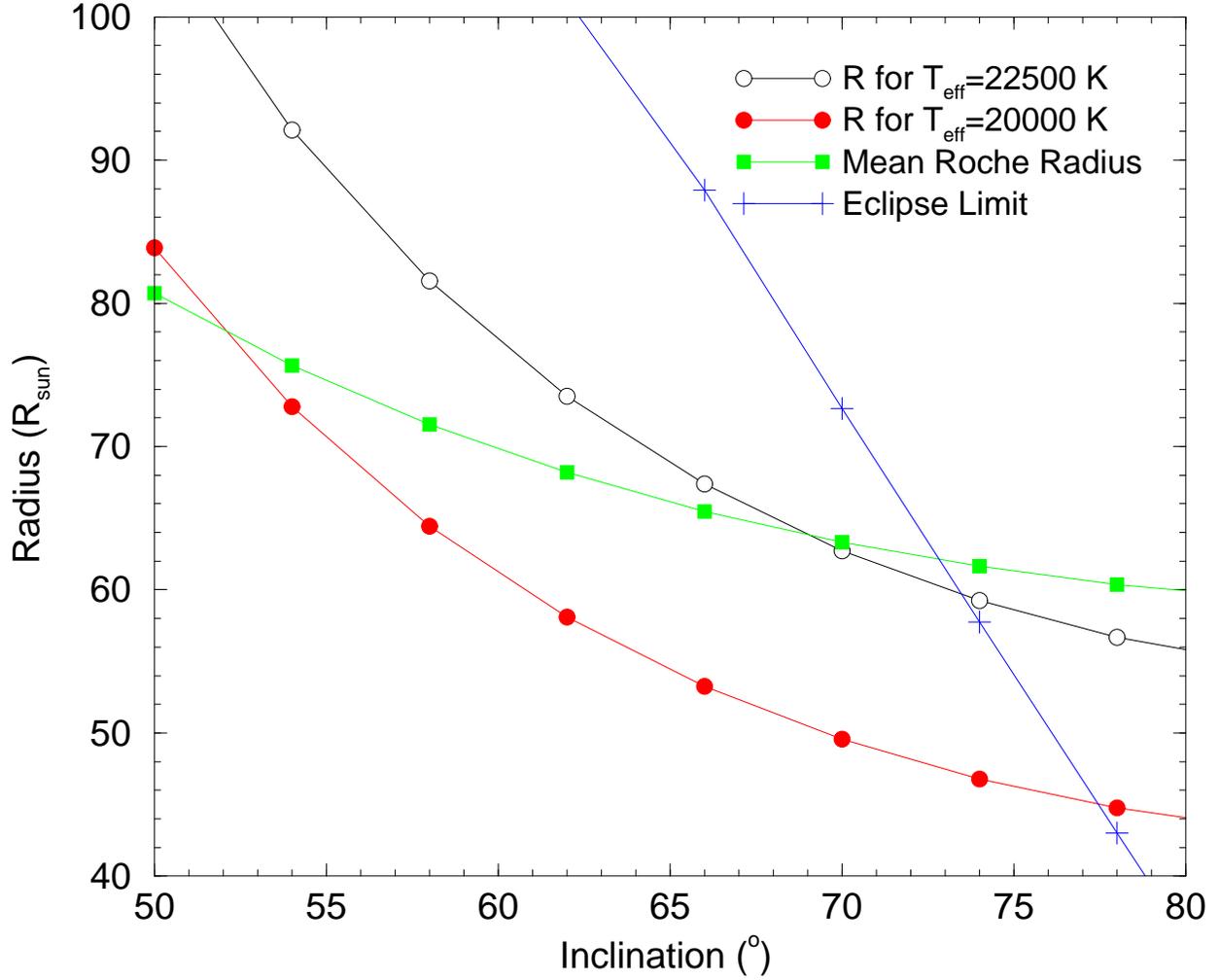}
\caption{Radius- orbital inclination constraints on WRAY 977. Absence of eclipse gives
an allowed region is left of the "Eclipse Limit" line; WRAY 977's mean radius is less 
than the "Mean Roche Radius"; mass-radius relations for upper and lower limits to 
$T_{eff}$ are plotted: the allowed region is between these two lines.}
%\label{Fig. 5}
\end{figure}

\begin{figure}
%\vspace{302pt}
%\includegraphics[width=84mm]{fig6.eps}
\plotone{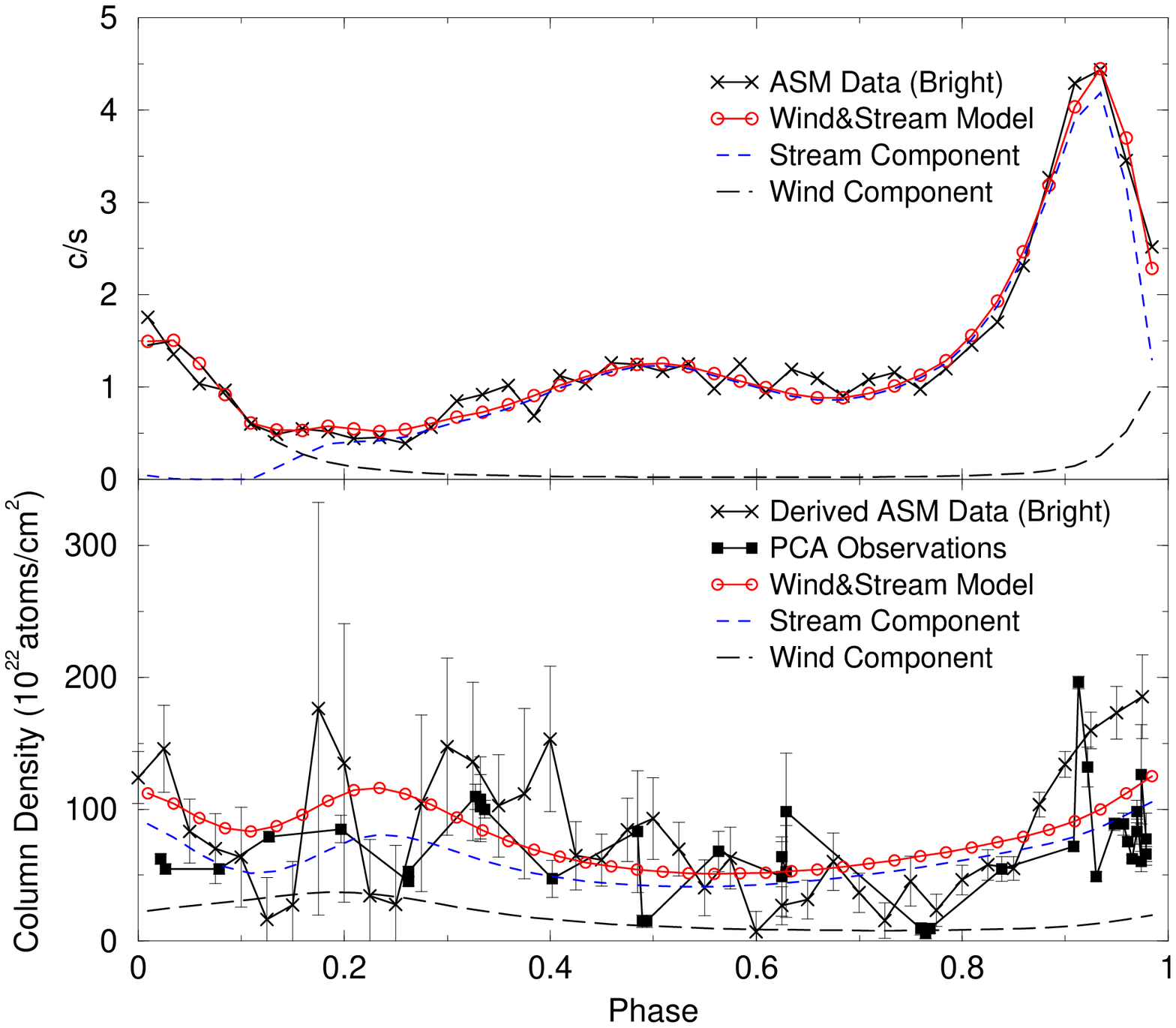}
\caption{Model fits to ASM lightcurve data (top panel) and column density data (bottom
panel) for bright level for $R_c=75R_{\odot}$, inclination 55$^\circ$. 
The contributions to light curve and column density from wind
and stream are shown separately, as well as the total.}
%\label{Fig. 6}
\end{figure}

%\bsp
\label{lastpage}


\begin{thebibliography}{99}
%\bibitem[\protect\citeauthoryear{Baird}{1981}]{b1} Baird S.R., 1981,
%ApJ, 245, 208
\bibitem[\protect\citeauthoryear{Auri\`ere}{1982}]{b1} Auri\`ere, M.  1982, A\&A,
    109, 301
\bibitem[\protect\citeauthoryear{Canizares et al.}{1978}]{b2} Canizares, C. R.,
    Grindlay, J. E., Hiltner, W. A., Liller, W., and
    McClintock, J. E.  1978, ApJ, 224, 39
%\bibitem[Castor, Abbott,\& Klein 1975]{cas75} Castor, J.,
	%Abbott, D. and Klein, R.  1975, \apj, 195, 157
\bibitem[\protect\citeauthoryear{Castor et al}{1975}]{b3} Castor, J.,
	Abbott, D. and Klein, R.  1975, ApJ, 195, 157
\bibitem[\protect\citeauthoryear{Djorgovski and King}{1984}]{b4} Djorgovski, S.,
    and King, I. R.  1984, ApJ, 277, L49
\bibitem[\protect\citeauthoryear{Haberl}{1991}]{b5}
Haberl, F. 1991, ApJ, 376, 245
\bibitem[\protect\citeauthoryear{Hagiwara and Zeppenfeld}{1986}]{b6} Hagiwara, K., and
    Zeppenfeld, D.  1986, Nucl.Phys., 274, 1
\bibitem[\protect\citeauthoryear{Harris and van den Bergh}{1984}]{b7} Harris, W. E.,
    and van den Bergh, S.  1984, AJ, 89, 1816
\bibitem[\protect\citeauthoryear{H\`enon}{1961}]{b8} H\'enon, M.  1961, Ann.d'Ap., 24, 369
\bibitem[\protect\citeauthoryear{Kaper et al.}{1995}]{b9}
Kaper, L., Lamers, H., Ruymaekers, E., van den Heuvel, E., \& 
Zuiderwijk, E. 1995, A\&A, 300, 446
\bibitem[\protect\citeauthoryear{King}{1966}]{b10} King, I. R.  1966, AJ, 71, 276
\bibitem[\protect\citeauthoryear{King}{1975}]{b11}  King, I. R.  1975, Dynamics of
    Stellar Systems, A. Hayli, Dordrecht: Reidel, 1975, 99
\bibitem[\protect\citeauthoryear{King}{1968}]{b12}  King, I. R., Hedemann, E.,
    Hodge, S. M., and White, R. E.  1968, AJ, 73, 456
\bibitem[\protect\citeauthoryear{Koh et al.}{1997}]{b13}
Koh, D., Bildsten, L., Chakrabarty, D. et al. 1997, ApJ, 479, 933
\bibitem[\protect\citeauthoryear{Kron et al.}{1984}]{b14} Kron, G. E., Hewitt, A. V.,
    and Wasserman, L. H.  1984, PASP, 96, 198
\bibitem[\protect\citeauthoryear{Leahy}{2002}]{b15} 
Leahy, D. 2002,  A\&A, 391, 219
\bibitem[\protect\citeauthoryear{Leahy}{1999}]{b16} 
Leahy, D., 1999, JRASC., 93, 33
\bibitem[\protect\citeauthoryear{Leahy and Creighton}{1993}]{b17}
Leahy, D., Creighton, J., 1993, MNRAS., 263, 314
\bibitem[\protect\citeauthoryear{Leahy}{1991}]{b18}
Leahy, D., 1991, MNRAS., 250, 310
\bibitem[\protect\citeauthoryear{Leahy \& Matsuoka}{1990}]{b19}
Leahy, D. \& Matsuoka, M. 1990, ApJ, 355, 627
\bibitem[\protect\citeauthoryear{Leahy \& Matsuoka}{1989a}]{b20}
Leahy, D.A., Matsuoka, M., Kawai, N. \& Makino, F. 1989, MNRAS, 236, 603
\bibitem[\protect\citeauthoryear{Leahy \& Matsuoka}{1989b}]{b21}
Leahy, D.A., Matsuoka, M., Kawai, N. \& Makino, F. 1989, MNRAS, 237, 269
\bibitem[\protect\citeauthoryear{Leahy}{1983}]{b22}
Leahy, D.A., Elsner, R., \& Weisskopf, M. 1983, ApJ, 272, 256
\bibitem[\protect\citeauthoryear{Levine et al.}{1996}]{b23}
Levine, A., Bradt, H., Cui, W.,et al. 1996, ApJ, 469, L33
\bibitem[\protect\citeauthoryear{Lynden-Bell and Wood}{1968}]{b24} Lynden-Bell, D.,
    and Wood, R.  1968, MNRAS, 138, 495
\bibitem[\protect\citeauthoryear{Mukherjee and Paul}{2004}]{b25} Mukherjee, U. and Paul, B. 
  2004, A\&A, 427, 567
\bibitem[\protect\citeauthoryear{Newell and O'Neil}{1978}]{b26} Newell, E. B.,
    and O'Neil, E. J.  1978, ApJS, 37, 27
\bibitem[\protect\citeauthoryear{Ortolani et al.}{1985}]{b27} Ortolani, S., Rosino, L.,
    and Sandage, A.  1985, AJ, 90, 473
\bibitem[\protect\citeauthoryear{Parkes et el.}{1980}]{b28}
Parkes, G., Mason, K., Murdin, P. \& Culhane, J. 1980, MNRAS, 191, 547
\bibitem[\protect\citeauthoryear{Peterson}{1976}]{b29} Peterson, C. J.  1976, AJ, 81, 617
\bibitem[\protect\citeauthoryear{Pravdo et al.}{1995}]{b30}
Pravdo, S., Day, C., Angelini, L., et al. 1996
ApJ, 454, 872
\bibitem[\protect\citeauthoryear{Saraswat et al.}{1996}]{b31}
Saraswat, P., Yoshida, A., Mihara, T. et al. 1996, ApJ, 463, 726
\bibitem[\protect\citeauthoryear{Sato et al.}{1986}]{b32}
Sato, N., Nagase, F., Kawai, N., et al. 1986
 ApJ, 304, 241
\bibitem[\protect\citeauthoryear{Shaller et al.}{1992}]{b33}
Shaller, G., Schaerer, D. \& Maeder, G. 1992, A\&AS, 96, 269
\bibitem[\protect\citeauthoryear{Spitzer}{1985}]{b34} Spitzer, L.  1985, Dynamics of
    Star Clusters, J. Goodman and P. Hut, Dordrecht: Reidel, 109
\bibitem[\protect\citeauthoryear{Stevens}{1988}]{b35}
Stevens, I. R. 1988 MNRAS, 232, 199
\bibitem[\protect\citeauthoryear{Watanabe et al.}{2003}]{b36}
Watanabe, S., et al. 2003, ApJ, 597, L37
\end{thebibliography}
\end{document}